\documentclass{ptapap}
\usepackage{amsmath}
\usepackage{amssymb}

\usepackage{color}
\usepackage{soul}

\author{Andrzej Pigulski}[IA]
\author{Piotr A.~Kołaczek-Szymański}[IA,UL]
\author{Marta Święch}[IA]
\author{Piotr Łojko}[IA]
\author{Kacper J.~Kowalski}[IA]
\affil[IA]{Uniwersytet Wrocławski, Instytut Astronomiczny, Kopernika 11, 51-622 Wrocław, Poland}
\affil[UL]{STAR, Université de Liège, B5c, Allée du 6 Août 19c, B-4000 Sart Tilman, Liège, Belgium}

\title{Magnetic blue large-amplitude pulsators}

\begin{document}

\maketitle

\begin{abstract}
With the discovery of blue large-amplitude pulsators (BLAPs), one of the most important questions to be answered was in which evolutionary channel(s) these radially pulsating stars could form. Located in the Hertzsprung-Russell diagram between hot subdwarfs and main sequence, BLAPs cannot be the products of single-star evolution. Therefore, virtually all scenarios explaining their origin assume binarity at some stage of evolution. One of these scenarios proposed that some BLAPs may have formed as a result of the merger of two low-mass white dwarfs or a white dwarf and a main-sequence star. The merger scenario may result in the formation of a magnetic field in the merger product, as is postulated for magnetic white dwarfs. Pulsations in a star with a strong magnetic field can manifest themselves in a way known from rapidly oscillating Ap stars, i.e.~in the form of evenly spaced multiplets (usually triplets) in the frequency spectrum. This phenomenon can be explained by the oblique pulsator model. We have shown that the three evenly spaced frequency modes discovered in the first BLAP, OGLE-BLAP-001, can be explained in this way. Similar behaviour has also been found in two other BLAPs. 
\end{abstract}
\section{Introduction}
About a decade ago, a new class of large-amplitude radially pulsating stars was discovered \citep{2017NatAs...1E.166P}, which came as quite a surprise in the world of variable stars. It seemed that, if only because of their large amplitudes, these stars should have been discovered long time ago. Due to their large photometric amplitudes and high effective temperatures, of the order of 30,000 K, these stars have been descriptively named blue large-amplitude pulsators (BLAPs). The reason why BLAPs remained unknown for so long time was, firstly, their extreme rarity and, secondly, the fact that they are intrinsically relatively faint stars. It took the analysis of the photometry of millions of stars to discover a larger number of them. In this context, the Optical Gravitational Lensing Experiment (OGLE) project was particularly useful. Thanks to data from this survey, 85\% of the about 200 currently known BLAPs have been discovered \citep{2017NatAs...1E.166P,2025ApJS..279...21P,2022...Brus_PhDT,2023AcA....73....1B,2023AcA....73..265B,2025AcA....75..223B}. The remaining BLAPs were discovered in data from other ground-based surveys \citep{2022MNRAS.511.4971M,2022MNRAS.513.2215R,2023NatAs...7..223L,2024MNRAS.529.1414C} and Transiting Exoplanet Survey Satellite (TESS) data \citep{2022A+A...663A..62P}.

Although the first star of this type to be discovered, OGLE-BLAP-001 = OGLE-GD-DSCT-0058, showed an equidistant triplet in its frequency spectrum \citep{2013AcA....63..379P}, it initially appeared that almost all stars of this type show only a single mode \citep{2020MNRAS.496.1105M}, presumably a fundamental radial one. We now know of several BLAPs in which more than one mode has been detected, raising hopes for the application of seismic modelling methods to them in the future. 

The most interesting problem associated with BLAPs seems to be the explanation of their origin, as their position in the Hertzsprung-Russell (H-R) diagram, between hot subdwarfs and massive main-sequence stars, indicates that they cannot be formed as the products of single-star evolution. Several scenarios for the formation of BLAPs have been proposed to date \citep{2018MNRAS.477L..30R,2018MNRAS.478.3871W,2018arXiv180907451C,2018MNRAS.481.3810B,2020MNRAS.492..232B,2020ApJ...903..100M,2021MNRAS.507..621B,2022A&A...668A.112X,2023ApJ...959...24Z,2024A&A...691A.103K}. All they assume that BLAPs form in binary systems. The masses of BLAPs predicted in these scenarios range from about 0.3\,M$_\odot$ to slightly more than 1\,M$_\odot$. Depending on the scenario and the current phase of evolution, models predict both a shortening and a lengthening of the pulsation periods of BLAPs. Given that these periods are very short (a typical period is about half an hour), this opens up an opportunity to use changes in these periods to determine the evolutionary scenario in which a particular BLAP was formed. Observations of BLAPs show both negative and positive rates of change in periods. This type of study of period changes also allows the discovery of the presence of BLAP companions, if they exist. Examples are HD 133729 \citep{2022A+A...663A..62P} and TMTS-BLAP-1 = ZGP-BLAP-01 \citep{2023NatAs...7..223L}. The observational properties of the entire BLAP group have recently been summarized by \cite{2025ApJS..279...21P}.

\section{Candidate magnetic BLAPs}
The presence of an equidistant triplet in the OGLE-BLAP-001 frequency spectrum required explanation. This was all the more so because, until recently, it was the only BLAP in which more than one frequency was observed. The explanation came from the fact that evenly-spaced frequency multiplets are observed in rapidly oscillating Ap (roAp) stars \citep{1982MNRAS.200..807K,1990ARA&A..28..607K}, a group of pulsating chemically peculiar magnetic stars. The pulsation properties in these stars are explained by the oblique pulsator model, i.e.~a star in which (i) there is a large-scale magnetic field, (ii) the axis of symmetry of the magnetic field is tilted relative to the axis of rotation, (iii) the axis of symmetry of the axisymmetric pulsation modes does not coincide with the axis of rotation. In general, the geometry of the magnetic field can be arbitrarily complex; what is important is that the presence of the magnetic field affects the pulsations in such a way that the observed amplitudes and phases of the pulsation modes depend on the phase of rotation. This produces equally-spaced multiplets in the frequency spectra. 

In roAp stars, we most often deal with non-radial modes, but modes which would be radial in the absence of magnetic field also occur. In the presence of magnetic field these modes are not purely radial as the presence of magnetic field distorts them \citep{1996ApJ...458..338D,2002A&A...391..235B}. Such modes are called distorted radial modes. An example of a distorted radial mode among roAp stars is the one observed in HR\,1217 \citep{2005MNRAS.358..651K}. Another example of an oblique radial pulsator is $\beta$ Cephei \citep{1997A&A...322..493T,2000ApJ...531L.143S}, the prototype of massive main-sequence pulsating stars in which a distorted radial mode is observed. Therefore, distorted radial modes in magnetic stars are known.

This prompted us to suggest \citep{2024A+A...691A.343P} that the equally-spaced triplet discovered in OGLE-BLAP-001 can be interpreted in terms of an oblique pulsator model. In the course of the analysis of Zwicky Transient Facility (ZTF) data for all known BLAPs, we found that another BLAP, ZGP-BLAP-08 \citep{2022MNRAS.511.4971M} = OGLE-BLAP-061 \citep{2023AcA....73....1B}, shows an equally-spaced triplet in its frequency spectrum. We therefore decided to analyse all available data of both stars. The two main goals of this analysis were (i) to verify if the modes are equally-spaced in frequency with high precision, (ii) examine the changes in periods in these stars. 

In the first step, we examined the changes in periods in these two stars by analysing each data set separately. Data sets that covered multiple seasons and were sufficiently long were divided into smaller fragments. This allowed us to examine period variability on timescales longer than approximately one year. The resulting O\,--\,C diagrams for OGLE-BLAP-001 and ZGP-BLAP-08 are shown in \cite{2024A+A...691A.343P}, their Figs.~1 and 6, respectively. They show that in both stars we are dealing with changes in period on a long time scale. Although these changes cannot be described by a constant rate of change in period, the long-term trend rather indicates a shortening of their pulsation periods. 

\begin{figure}
\centering
\includegraphics[width=0.9\textwidth]{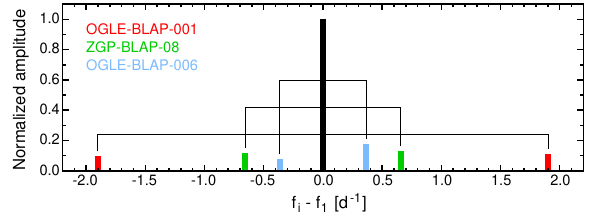}
\caption{A schematic view of the frequency triplets observed in three candidate magnetic BLAPs, OGLE-BLAP-001 (red), ZGP-BLAP-08 (green), and OGLE-BLAP-006 (blue). The $f_i$ stands for both $f_{\rm left}$ and $f_{\rm right}$ from Table \ref{tab:freq}, $f_1$ is the central frequency of a triplet.}
\label{FS-MBLAPs}
\end{figure}

In the next step, we combined all good-quality data to perform a final analysis aimed at detecting low-amplitude components in the frequency spectra. In order to improve their visibility, we first corrected the observation times for period changes obtained in the previous step. Such a correction improves the visibility of modes only if they change their period similarly to the dominant mode. This may not always be the case, but this can be verified by analysing the data with and without this correction. Additionally, since the photometric observations were made in different bands, we normalized all data subsets to the amplitudes of the dominant mode. The observations, corrected for period changes and normalized to the amplitude of the main mode, were then subjected to a final analysis. The result of this analysis was the confirmation of the presence of three equidistant frequencies in the frequency spectra of both stars. A schematic view of their frequency spectra is shown in Fig.\,\ref{FS-MBLAPs}. Since the light curves are non-sinusoidal, frequency spectra contained also harmonics of the dominant mode and low-order combination frequencies of the dominant frequencies and the side components. The dominant frequencies, $f_1$, side frequencies, $f_{\rm right}$ and $f_{\rm left}$, and other characteristics of the frequency triplets are summarized in Table \ref{tab:freq}. The numbers clearly indicate that within the errors the side frequencies are evenly separated from the central frequency.

\begin{table}[h]
\caption{Characteristics of frequency triplets in the frequency spectra of three candidate magnetic BLAPs.}
\label{tab:freq}
\begin{center}
\small
\begin{tabular}{c|ccc}
\hline
\noalign{\smallskip}
Quantity | Star:  & OGLE-BLAP-001 & ZGP-BLAP-08 & OGLE-BLAP-006 \\ 
\noalign{\smallskip}
\hline \hline
\noalign{\smallskip}
Main frequency, $f_1$ [d$^{-1}$] & 50.9645803(10) & 40.9822311(26) & 37.8797546(6)\\ 
$\Delta_1=f_{\rm right}-f_1$ [d$^{-1}$] &  1.902964(11) & 0.657407(21) & 0.365387(05) \\
$\Delta_2=f_1-f_{\rm left}$ [d$^{-1}$] & 1.902928(13) & 0.657474(24)&  0.365410(10)\\
$2(\Delta_1 - \Delta_2)/(\Delta_1 + \Delta_2)$ & 4\,$\times$\,10$^{-5}$ & 1.0\,$\times$\,10$^{-4}$ & 6\,$\times$\,10$^{-5}$ \\
Rotation period [d] & 0.5255 & 1.5211 & 2.7367 \\
$A_{\rm left}/A_1$ & 0.093(17) & 0.118(24) & 0.078(5) \\
$A_{\rm right}/A_1$ & 0.112(17)& 0.134(24) & 0.178(5) \\
\noalign{\smallskip}
\hline
\end{tabular}
\end{center}
\end{table}

Confirmation of the same spacing of side frequencies in OGLE-BLAP-001 and the discovery of a similar structure in the frequency spectrum of ZGP-BLAP-08 indicates that, as in the roAp stars, we may be dealing with magnetic stars. Importantly, this is consistent with two scenarios for the formation of BLAPs as merger products, as the coalescence of two stars leads to an enhancement in the magnetic field \citep{2019Natur.574..211S}. It is therefore possible that these two BLAPs are the products of mergers, either of a white dwarf and a main-sequence star \citep{2023ApJ...959...24Z} or of two low-mass white dwarfs \citep{2024A&A...691A.103K}. Although these are rarely realized evolutionary channels, the estimated incidence of BLAPs is consistent with observations.

Shortly after the publication of our results on two magnetic BLAP candidates, \cite{2025AJ....169..261K} published a detailed photometric analysis of another BLAP, OGLE-BLAP-006. Their analysis was based primarily on excellent photometry from the ground-based Korea Microlensing Telescope Network (KMTNet) project carried out in the years 2016\,--\,2023, supplemented by the OGLE data. They found three independent pulsation modes and rather complex period variations of the dominant mode. These variations were interpreted by them in terms of the light-travel time effect (LTTE) in a triple system, superimposed on a long-term evolutionary changes with a negative constant rate. The derived orbital periods of the two hypothetical companions were equal to 13 and 17 years.

Three equidistant frequencies appear in the OGLE-BLAP-006 frequency spectrum. \cite{2025AJ....169..261K} interpret them as two independent frequencies, $f_1$ and $f_2$, and a combination frequency $f_3 = 2f_1 - f_2$, arguing that the amplitude of $f_3$ is significantly smaller than $f_2$. However, in an equidistant triplet it is always possible to interpret one of the components as a combination frequency. From the point of view of the theory of oblique rotators, asymmetries in amplitudes of side frequencies are allowed and depend primarily on the geometry of the magnetic field. An example is provided by \cite{1996MNRAS.281..883K}. Therefore, we believe that interpreting all three frequencies ($f_1$, $f_2$ and $f_3$ from their Table 1) as independent frequencies is entirely justified. Consequently, we consider OGLE-BLAP-006 to be the third candidate for a magnetic BLAP, and the relevant data for its frequency triplet are shown in Table \ref{tab:freq} and in Fig.\,\ref{FS-MBLAPs}. 

\section{Conclusions}
One may wonder whether the explanation of equidistant frequency triplets by means of the oblique pulsator model is the only possibility. In general, no, because other mechanisms can also produce equidistant triplets. First, such a triplet can also be produced when a pulsating star resides in a binary system. The best example is another BLAP, HD\,133729 \citep{2022A+A...663A..62P}, in which such a triplet appears and is a manifestation of the binarity of this BLAP. However, in HD\,133729, the orbital period is 23.08~d, which gives a separation between the components of the triplet equal to 0.0433\,d$^{-1}$, much smaller than for the three BLAPs discussed here (Table \ref{tab:freq}). If the splittings in the three BLAPs were interpreted as orbital frequencies, we would obtain unrealistically high masses for the components. In addition, strong proximity effects would be observed, which are not visible in their photometry. This falsifies this possibility. The next possibility is the model discussed by \cite{1985A&A...151..315M}, which assumes spots on the stellar surface. However, this model suffers from physically questionable assumptions and also predicts changes with the frequency of rotation, which are not visible in the photometry of the three BLAPs. The final possibility is rotational splitting. This, however, does not occur for radial modes and, moreover, rarely produces perfectly equidistant frequencies at the level of 10$^{-5}$ of the central frequency. In summary, the oblique pulsator model remains the only acceptable explanation for evenly spaced frequencies in the three candidates for magnetic BLAPs. 

We currently know of three BLAPs showing equidistant triplets in their frequency spectra. These modes are easily explained in the oblique pulsator model, which makes these stars excellent candidates for magnetic BLAPs. Going further, it is very likely that these stars were formed as products of mergers, which naturally explains the presence of magnetic fields in them. Of course, the presence of strong magnetic fields can be verified by spectropolarimetric observations, although this will not be easy because these three BLAPs are faint objects ($V$ magnitudes in the range of 17.4\,--\,18.4 mag). As Table \ref{tab:freq} shows, the side components have amplitudes of about 10\% of the amplitude of the main component, so it can be expected that detailed observations of other BLAPs will allow the discovery of further candidates for magnetic BLAPs. The important thing is that BLAPs constitute yet another group in which it is possible to study the relationships between pulsations and the magnetic field and to consider evolutionary scenarios involving stellar mergers.

\acknowledgements{This work was supported by the National Science Centre, Poland, grant no. 2022/45/B/ST9/03862.}

\bibliographystyle{ptapap}
\bibliography{Pigulski}
\end{document}